\documentclass[a4paper,english,aps,prl,twocolumn,showpacs,superscriptaddress,groupedaddress]{revtex4}
\usepackage[T1]{fontenc}
\usepackage[latin9]{inputenc}
\usepackage{babel}
\usepackage{array}
\usepackage{graphicx}
\usepackage[unicode=true,pdfusetitle,
 bookmarks=true,bookmarksnumbered=false,bookmarksopen=false,
 breaklinks=false,pdfborder={0 0 1},backref=false,colorlinks=false]
 {hyperref}

\makeatletter

\pdfpageheight\paperheight
\pdfpagewidth\paperwidth

\DeclareRobustCommand{\greektext}{%
  \fontencoding{LGR}\selectfont\def\encodingdefault{LGR}}
\DeclareRobustCommand{\textgreek}[1]{\leavevmode{\greektext #1}}
\DeclareFontEncoding{LGR}{}{}
\DeclareTextSymbol{\~}{LGR}{126}
\providecommand{\tabularnewline}{\\}

\@ifundefined{textcolor}{}
{%
 \definecolor{BLACK}{gray}{0}
 \definecolor{WHITE}{gray}{1}
 \definecolor{RED}{rgb}{1,0,0}
 \definecolor{GREEN}{rgb}{0,1,0}
 \definecolor{BLUE}{rgb}{0,0,1}
 \definecolor{CYAN}{cmyk}{1,0,0,0}
 \definecolor{MAGENTA}{cmyk}{0,1,0,0}
 \definecolor{YELLOW}{cmyk}{0,0,1,0}
}

\usepackage{diagbox}
\usepackage{makecell}
\usepackage{amssymb}

\makeatother

\begin{document}

\title{}

\title{Observation of OAM sidebands due to optical reflection}

\author{W. Löffler}

\email{loeffler@physics.leidenuniv.nl}

\affiliation{Huygens Laboratory, Leiden University, P.O. Box 9504, 2300 RA Leiden,
The Netherlands}

\author{Andrea Aiello}

\affiliation{Max Planck Institute for the Science of Light, Günther-Scharowsky-Straße
1/Bldg. 24, 91058 Erlangen, Germany}

\affiliation{Institute for Optics, Information and Photonics, Universität Erlangen-Nürnberg,
Staudtstr. 7/B2, 91058 Erlangen, Germany}

\author{J. P. Woerdman}

\affiliation{Huygens Laboratory, Leiden University, P.O. Box 9504, 2300 RA Leiden,
The Netherlands}
\begin{abstract}

We investigate how the orbital angular momentum (OAM) of a paraxial
light beam is affected upon reflection at a planar interface. Theoretically,
the unavoidable angular spread of the (paraxial) beam leads to OAM
sidebands which are found to be already significant for modest beam
spread (0.05). In analogy to the polarization Fresnel coefficients
we develop a theory based upon spatial Fresnel coefficients; this
allows straightforward prediction of the strength of the sidebands.
We confirm this by experiment.
\end{abstract}
\maketitle

A light beam, either classical or quantum, possesses spatially transverse
degrees of freedom. A very popular example is the Orbital Angular
Momentum (OAM) of light \cite{allen1992,allen1999}; this has important
applications in quantum communication \cite{mair2001,molinaterriza2007,dada2011}.
The key advantage of OAM for quantum communication is its high dimensionality
\cite{pors2011}; this allows a single photon to carry much more \cite{bechmannpasquinucci2000,gibson2004}
than the two bits of information (qubit) enabled by the polarization
degree of freedom. In this Letter we investigate, theoretically and
experimentally, the effect of optical reflection at a planar interface
on the OAM state of a beam. This is a relevant issue since reflection
is the simplest possible optical operation, and is often unavoidable
in experiments and applications. An analogous issue is well known
for the case of light\textquoteright{}s polarization; this is generally
strongly affected by Fresnel reflection (apart from special cases
\footnote{Only the eigenstates, s or p polarization, remain pure upon reflection,
albeit attenuated%
}). What happens in the OAM case? Everyday experience tells us that
image distortion does not occur if we use a planar mirror; this suggests
that the spatial state of the input beam (and thus also the OAM spectrum)
should be well preserved upon reflection. 

As we will see, the OAM state is indeed conserved if we use an ideal
mirror, which we define as a (planar) mirror with infinite dielectric
contrast. However, a practical mirror has a finite dielectric contrast;
think for instance of a single dielectric interface, or a multi-layer
dielectric mirror, or a metal mirror. In this case wave optics leads
to diffractive corrections upon Snell\textquoteright{}s Reflection
Law such as the Goos-Hänchen \cite{goos1947} (GH) and Imbert-Fedorov
\cite{fedorov1955,imbert1972} (IF) shifts: the reflected beam is
shifted relative to the geometrical-optics reflected ray \cite{bliokhprl2006,hosten2008,aiellobp2008,schwefel2008,gilles2002,bliokhpre2007}.
A natural theoretical description is furnished by spatial Fresnel
coefficients that act upon the transverse modes of the incident light
beam, analogous to the conventional polarization Fresnel coefficients
that act upon the polarization modes. We find that the reflection-induced
modification of the OAM state depends on the angular spread of the
beam; this leads to OAM sidebands, even in the paraxial approximation.
This is confirmed by our experimental results.

\begin{figure}
\includegraphics[width=1\columnwidth]{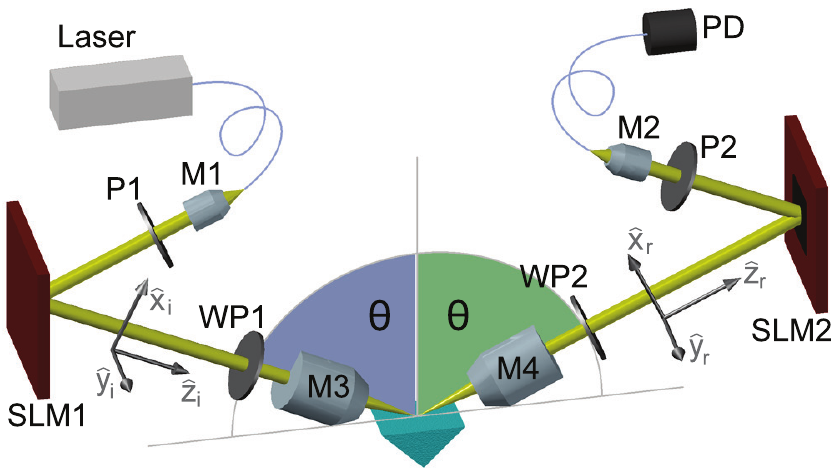}\caption{\label{fig:experiment}Experimental setup: Light from a single mode
fiber pigtailed laser is collimated (M1), polarized (P1), and then
modulated using a phase-only spatial light modulator (SLM1) to prepare
a certain mode. This light is then reflected at the interface, and
the reflected light is analyzed using a combination of SLM2 and a
single mode fiber connected to a photodiode (PD). Microscopy objectives
M3 \& M4 (10x, 0.25 NA) can be introduced to change the beam focussing
($\theta_{0}$), $\lambda/2$ waveplates WP1\&2 modulate the polarization.}
\end{figure}

Our work connects with several papers that report on OAM spectral
broadening due to misalignment of the reference frame with respect
to the OAM beam \cite{gibson2004,vasnetsov2005,zambrini2006}. In
an experiment this type of broadening can obviously be removed by
proper re-adjustment of the set-up to compensate the original misalignment.
However, in our case we deal with intrinsic displacements (due to
GH and IF shifts); this case is essentially different since, as we
will see, intrinsic displacement depends on the spatial (OAM) mode
and can thus not be cancelled simply by optical adjustment if we deal
with a superposition of spatial modes as input state. We aim here
to quantify the corresponding OAM spectral broadening. Our work also
connects with that of Okuda and Sasada \cite{okuda2006,okuda2008}
who study giant (nonperturbative) deformation of a OAM mode due to
total internal reflection (TIR) for incidence very close to the critical
angle, where a singularity occurs. In this regime beam shifts (GH
and IF) and thus the OAM mode spectrum are ill-defined concepts \cite{okuda2008};
we do not consider this singular TIR case in the present Letter.

\textbf{}

Theoretically, we describe the reflection process in terms of a scattering
operator $\hat{S}=\sum_{\lambda}\hat{P}_{\lambda}\otimes\hat{M}_{\lambda}$,
where $\hat{P}_{\lambda}$ acts on the polarization state $|\lambda\rangle$
($\lambda=1$ and $\lambda=2$ correspond to $p$ (in plane) and $s$
(out of plane) polarization, respectively), and $\hat{M}_{\lambda}$
on the spatial state $|\psi\rangle$. Here we have adopted a quantum
notation for the sake of clarity. We restrict ourselves to a paraxial
light field, in this case polarization and spatial degree of freedom
factorize: $|in\rangle=|\lambda\rangle|\psi\rangle$. Upon reflection,
$\hat{S}$ mixes the polarization and spatial part; thus after scattering,
it is not possible to write the state as before in a product of polarization
times spatial state \cite{bliokhpre2007,aiello2004b}; further this
enables the link between beam shifts and weak measurements \cite{dennis2012,hosten2008}.

In more detail, we discuss the incoming field $\mathbf{E}_{i}(x_{i},y_{i},z_{i},t)=\mathrm{Re}[\mathbf{A}_{i}(x_{i},y_{i},z_{i})\,\exp(-i\omega t)]$
in terms of its \emph{analytic signal }\cite{bliokhvortex2009,merano2010,aiello2012}
$\mathbf{A}_{i}(x_{i},y_{i},z_{i})=\sum_{\lambda}a_{\lambda}\hat{\mathbf{e}}'_{\lambda}\psi(x_{i},y_{i},z_{i})$
where $\psi$ describes the spatial shape of the beam (which we keep
arbitrary at this point), $\hat{\mathbf{e}}'_{1}=\hat{\mathbf{x}}_{i}$
and $\hat{\mathbf{e}}'_{2}=\hat{\mathbf{y}}_{i}$, are the incoming-beam
unit vectors, and $a_{1,2}$ are the polarization coefficients. All
coordinates are expressed in units of $k$, and hence dimensionless.
The reflected field can be written as 
\begin{equation}
\mathbf{A}(x,y,z)=\sum_{\lambda}a_{\lambda}r_{\lambda}(\theta)\psi(-x+X_{\lambda},y-Y_{\lambda},z)\hat{\mathbf{e}}_{\lambda}\label{eq:aa29}
\end{equation}
 where $\hat{\mathbf{e}}_{1}=\hat{\mathbf{x}}_{r}$ and $\hat{\mathbf{e}}_{2}=\hat{\mathbf{y}}_{r}$
are the unit vectors in the reflected-beam coordinate system (Fig.~\ref{fig:experiment}).
The reflected field $\mathbf{A}$ depends on the Fresnel reflection
coefficients $r_{\lambda}$ and the four complex beam shifts $X_{\lambda}$
and $Y_{\lambda}$, whose real (imaginary) part corresponds to spatial
(angular) longitudinal Goos-Hänchen \cite{goos1947} and transverse
Imbert-Fedorov \cite{imbert1972,bliokhprl2006} shifts, respectively
\cite{merano2010}: $X_{\lambda}=-i\,\partial_{\theta}\left[\ln r_{\lambda}(\theta)\right]$
and $Y_{1}=-i\frac{a_{2}}{a_{1}}\left(1+\frac{r_{2}}{r_{1}}\right)\cot\theta$,
$Y_{2}=i\frac{a_{1}}{a_{2}}\left(1+\frac{r_{1}}{r_{2}}\right)\cot\theta$.
By analyzing $X_{\lambda}$ and $Y_{\lambda}$ it can easily be seen
that only an infinite refractive index contrast makes them disappear,
so only at reflection from such a perfect mirror, the reflected mode
$\mathbf{A}(x,y,z)$ is not perturbed. Since the combined shift $\mathbf{R}_{\lambda}=\left(X_{\lambda},Y_{\lambda}\right)$
is supposedly small, we can Taylor-expand the shifted function $\psi$
to find deviations from geometrical optics reflection. We obtain for
the spatial part with $\mathbf{R}=\left(x,y\right)$:

\begin{eqnarray}
\langle x,y,z|\hat{M}_{\lambda}|\psi\rangle & = & \psi(-x+X_{\lambda},y-Y_{\lambda},z)\nonumber \\
 & \simeq & \psi(-x,y,z)+\mathbf{R}_{\lambda}\cdot\frac{\partial}{\partial\mathbf{R}}\psi(-x,y,z)
\end{eqnarray}

\begin{figure}
\includegraphics[width=1\columnwidth]{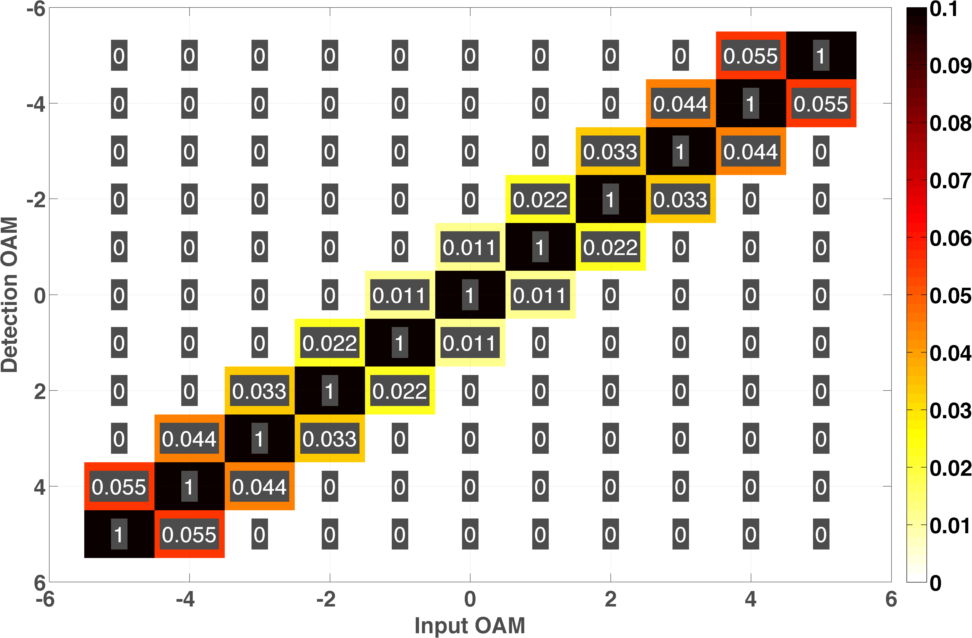}\caption{\label{fig:resth1}Calculated SFC intensity $C_{\ell,\ell'}^{p}=|c_{\ell,\ell'}^{p}|^{2}$
for the case of $p$-polarized OAM modes externally reflected at a
dielectric interface ($n=1.52,\,\theta=70^{\circ},\,\theta_{0}=0.05$).
Input ($\ell$) and output ($\ell'$) OAM is shown on the horizontal
and vertical axis, respectively (radial mode number $p=0$). }
\end{figure}

Now, we specialize to a specific mode basis, and expand the spatial
part of the input field in terms of the Laguerre-Gauss modes $\phi_{\ell}^{p}$
as $|\psi\rangle=\sum_{\ell,p}|\phi_{\ell}^{p}\rangle\langle\phi_{\ell}^{p}|\psi\rangle$.
Our goal is to predict the reflected state, from which we can derive
the spatial mode scattering matrix $\hat{M}_{\lambda}$, which has
as elements the spatial Fresnel coefficients (SFC)

\begin{equation}
c_{\ell,p,\ell',p'}^{\lambda}\equiv\langle\phi_{\ell'}^{p'}|\hat{M}_{\lambda}|\phi_{\ell}^{p}\rangle\label{eq:sfcaa12-1}
\end{equation}

To obtain these coefficients, we use two properties of LG modes: (i)
$\phi_{\ell}^{p}(-x,y,z)=(-1)^{\ell}\phi_{-\ell}^{p}(x,y,z)$ and
(ii) the known spatial derivatives of LG modes, $\frac{\partial\phi_{\ell}^{p}}{\partial x}$
and $\frac{\partial\phi_{\ell}^{p}}{\partial y}$, up to first order
in $\theta_{0}\,(=\lambda/\pi\omega_{0})$, the mode half-opening
angle. We find the following non-zero coefficients $c_{\ell,p,\ell',p'}^{\lambda}$:\renewcommand{\arraystretch}{1.4}

\begin{tabular}{|ccc|c|c|}
\hline 
\diagbox{$p'$}{$\ell'(\ell\gtrless 0)$} & \vrule width 1.5pt & $-\ell\mp1$ & $-\ell$ & $-\ell\pm1$\tabularnewline
\hline 
\Xhline{1.5pt}$p-1$ & \vrule width 1.5pt & $\pm Z_{\lambda}^{\pm}\sqrt{p}$ & $0$ & $0$\tabularnewline
\hline 
$p$ & \vrule width 1.5pt & $\pm Z_{\lambda}^{\pm}\sqrt{|\ell|+p+1}$ & $(-1)^{\ell}$ & $\mp Z_{\lambda}^{\mp}\sqrt{|\ell|+p}$\tabularnewline
\hline 
$p+1$ & \vrule width 1.5pt & $0$ & $0$ & $\mp Z_{\lambda}^{\mp}\sqrt{p+1}$\tabularnewline
\hline 
\end{tabular}

\begin{tabular}{|ccc|c|c|}
\hline 
\diagbox{$p'$}{$\ell'\,(\ell=0)$} & \vrule width 1.5pt & $-1$ & $0$ & $+1$\tabularnewline
\hline 
\Xhline{1.5pt}$p-1$ & \vrule width 1.5pt & $Z_{\lambda}^{+}\sqrt{p}$ & $0$ & $0$\tabularnewline
\hline 
$p$ & \vrule width 1.5pt & $Z_{\lambda}^{+}\sqrt{p+1}$ & $(-1)^{\ell}$ & $-Z_{\lambda}^{-}\sqrt{p}$\tabularnewline
\hline 
$p+1$ & \vrule width 1.5pt & $0$ & $0$ & $-Z_{\lambda}^{-}\sqrt{p+1}$\tabularnewline
\hline 
\end{tabular}

\vspace{-1.1cm}
\begin{equation}
\label{eq:aa39}
\end{equation}

Here, we can elegantly combine all occurring shifts, i.e., the longitudinal
and transverse, in each case the spatial and angular variant, in a
single complex number $Z_{\lambda}^{\pm}=\frac{\theta_{0}}{2^{3/2}}(-1)^{\ell}\left(X_{\lambda}\pm i\, Y_{\lambda}\right)$.
This is specific to Laguerre-Gauss modes. Fig.~\ref{fig:resth1}
shows exemplary the SFC intensities $C_{\ell,\ell'}^{p}=|c_{\ell,\ell'}^{p}|^{2}$
for reflection of p-polarized purely azimuthal ($p=0$) LG modes with
$\theta_{0}=0.05$ at an angle of incidence of $\theta=70^{\circ}$
under external reflection at an air-BK7 interface. We find that reflection
described by the SFCs induces a transformation of a pure $\{\ell\}$
mode into a superposition of $\{-\ell-1,-\ell,-\ell+1\}$ modes, where
the minus sign stems simply from OAM reversal upon reflection. The
coupling strength to the OAM sideband modes $\{-\ell-1,-\ell+1\}$
is proportional to $\theta_{0}^{2}$ and depends linearly on $\ell$.
In our case of pure azimuthal modes, the coupling strength is simply
proportional to $\left(\theta_{0}\sqrt{|\ell|}\right)^{2}$; here
we recognize the effective mode opening angle of which is proportional
to $\theta_{0}\sqrt{|\ell|}$. We also see that the mode coupling
is governed by intrinsic displacement induced by beam shifts $X_{\lambda}$
and $Y_{\lambda}$. In the particular case of external reflection
at a dielectric, only angular beam shifts occur \cite{bliokhprl2006,aiellobp2008};
this can be seen by analyzing $X_{\lambda}$ and $Y_{\lambda}$: for
linear $s$ or $p$ polarization, $Y_{\lambda}$ vanishes, and $X_{\lambda}$
is purely imaginary, this corresponds to an angular shift within the
plane of incidence. 

\begin{figure}
\includegraphics[width=1\columnwidth]{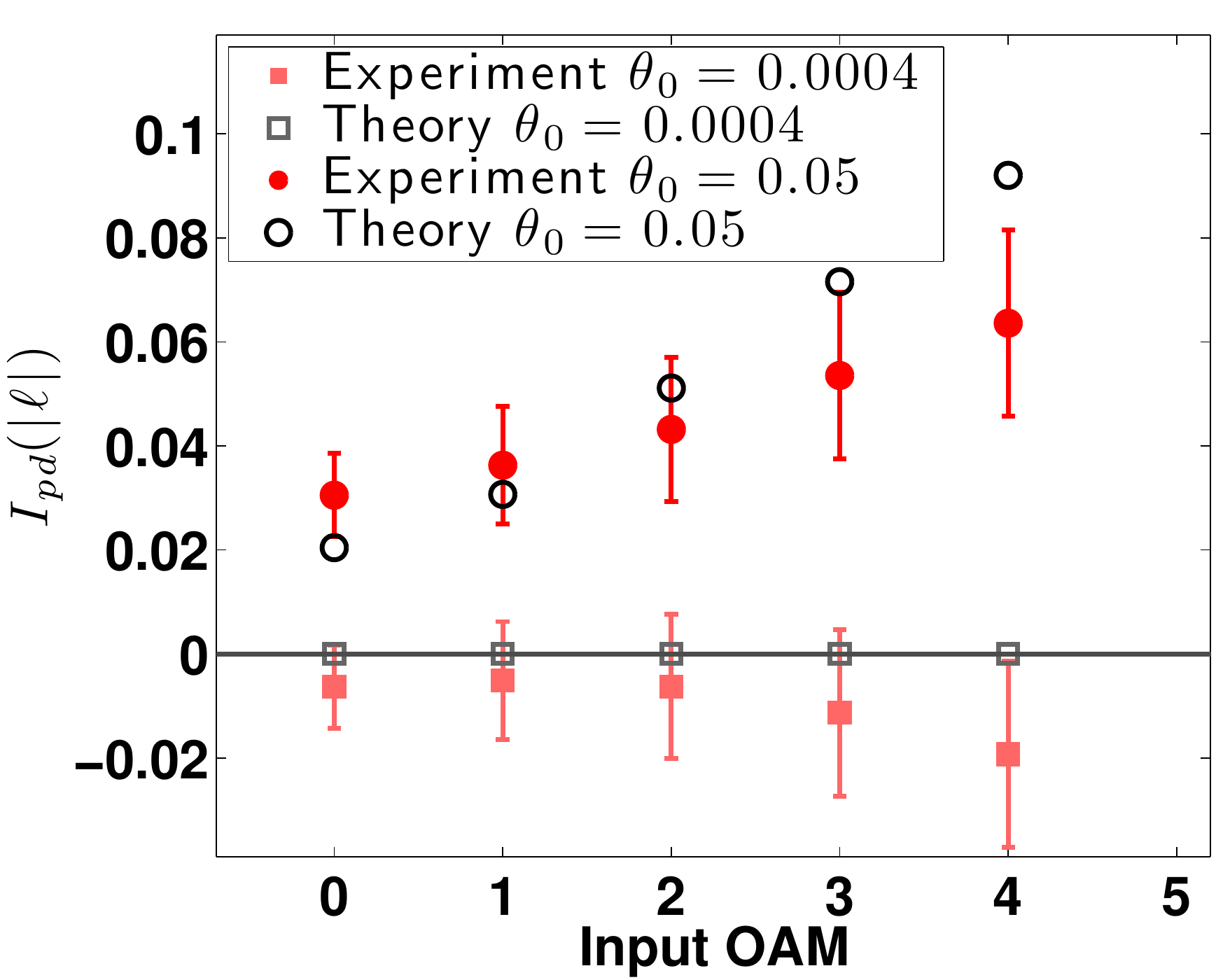}\caption{\label{fig:resexp}Experimental and theoretical polarization-differential
SFC sideband intensity $I_{pd}(\ell)$ as a function of input OAM
($\ell$) for $p=0$, for the case of external reflection by an air-glass
interface, at an angle of incidence $\theta=70^{\circ}$. Error bars
are estimated from multiple measurement runs to take mechanical drifts
and misalignment into account. }
\end{figure}

The discussion above was for the case of a single OAM input state;
however it is straightforward to extend this to an OAM superposition
input state. The mode coupling by reflection is a coherent scattering
process so that a superposition can be handled by decomposition into
its constituent OAM modes. 

As a demonstration experiment, we investigate the case displayed in
Fig.~\ref{fig:resth1}, i.e. external reflection at an air - glass
interface. As shown in Fig.~\ref{fig:experiment} , light from a
fiber pigtailed 635~nm laser is collimated using a 20x microscopy
objective (beam waist $\omega_{0}=530$~\textgreek{m}m, this corresponds
to $\theta_{0}=0.0004$). The light is sent to a spatial light modulator
(SLM1) (10~\textgreek{m}m pixel size, 3~mrad blaze angle), which
imprints the desired azimuthal phase to produce the input OAM spectrum
($\ell$; $p=0$). The beam is then focused, externally reflected
at the hypothenuse plane of a standard BK7 ($n=1.52$) glass prism,
and then recollimated to a beam waist of $\omega_{0}=530$~\textgreek{m}m.
For this telescope configuration we use underfilled 10x microscopy
objectives (to minimize aberrations) M3 and M4 (0.25 NA) to obtain
$\theta_{0}=0.05$ inside the telescope, which is well within the
paraxial approximation. The reflected beam $|out\rangle$ is sent
to SLM2 (phase profile $\ell'$) which is followed by microscopy objective
M4 and a single-mode fiber; thus we project on a pure OAM mode ($\ell'$)
\footnote{The reflected-beam axis is adjusted to be the axis of the reflected
$s$-polarized fundamental Gaussian beam.%
}. The fiber is connected to a photodiode (PD); its photocurrent is
proportional to $C_{\ell,\ell'}$. By scanning the OAM of the input
($\ell$) and output ($\ell'$) modes, we map the matrix containing
the spatial Fresnel coefficients. From these data we deduce the relative
intensity in the OAM sidebands compared to the total reflected intensity 

\begin{equation}
I_{rel}^{\lambda}(\ell)=\frac{C_{\ell,-\ell-1}^{\lambda}+C_{\ell,-\ell+1}^{\lambda}}{\sum_{\ell'}C_{\ell,\ell'}^{\lambda}}\label{eq:sfcnorm}
\end{equation}
In order to improve the signal-to-noise ratio we use polarization
modulation ($s$ versus $p$) with a half-wave plates (WP1 and WP2)
before and after reflection from the interface; this enables polarization-differential
measurement. The experiment thus yields the polarization-differential
$I_{pd}(\ell)=I_{rel}^{p}(\ell)-I_{rel}^{s}(\ell)$, which is plotted
in Fig.~\ref{fig:resexp} versus the input $\ell$, at a fixed angle
of incidence ($\theta_{0}=70^{\circ}$). The data for $\theta_{0}=0.0004$
have been obtained without the telescope. In this case Eq.~\ref{eq:aa39}
predicts polarization-differential sidebands $<10^{-5}$ in the experimentally
addressed $\ell$ range, which is much smaller than (but consistent
with) the experimental accuracy. For $\theta_{0}=0.05$ the mode coupling
is increased, in reasonable agreement with theory %
\footnote{Experimental errors are ascribed to drift and residual misalignment;
apart from proper adjustment of the setup and the holograms we had
to align very precisely the orientation and position of the microscopy
objectives, and stability is of extreme importance.%
}. From experiments, as well as from numerical simulation, we observe
that mode coupling to $-\ell\pm2$ modes is at least an order of magnitude
smaller than the coupling to $-\ell\pm1$ %
\footnote{The Brewster angle is at $\theta_{B}=56.7^{\circ}$. In Fig.3 we show
results for an angle of incidence of $\theta=70^{\circ}$, this is
$5\cdot\theta_{0}$ separated from the Brewster resonance \cite{merano2009}
at $\theta_{B}=56.7^{\circ}$, thus we operate truly in the perturbative
regime.%
}.

\begin{figure}
\includegraphics[width=1\columnwidth]{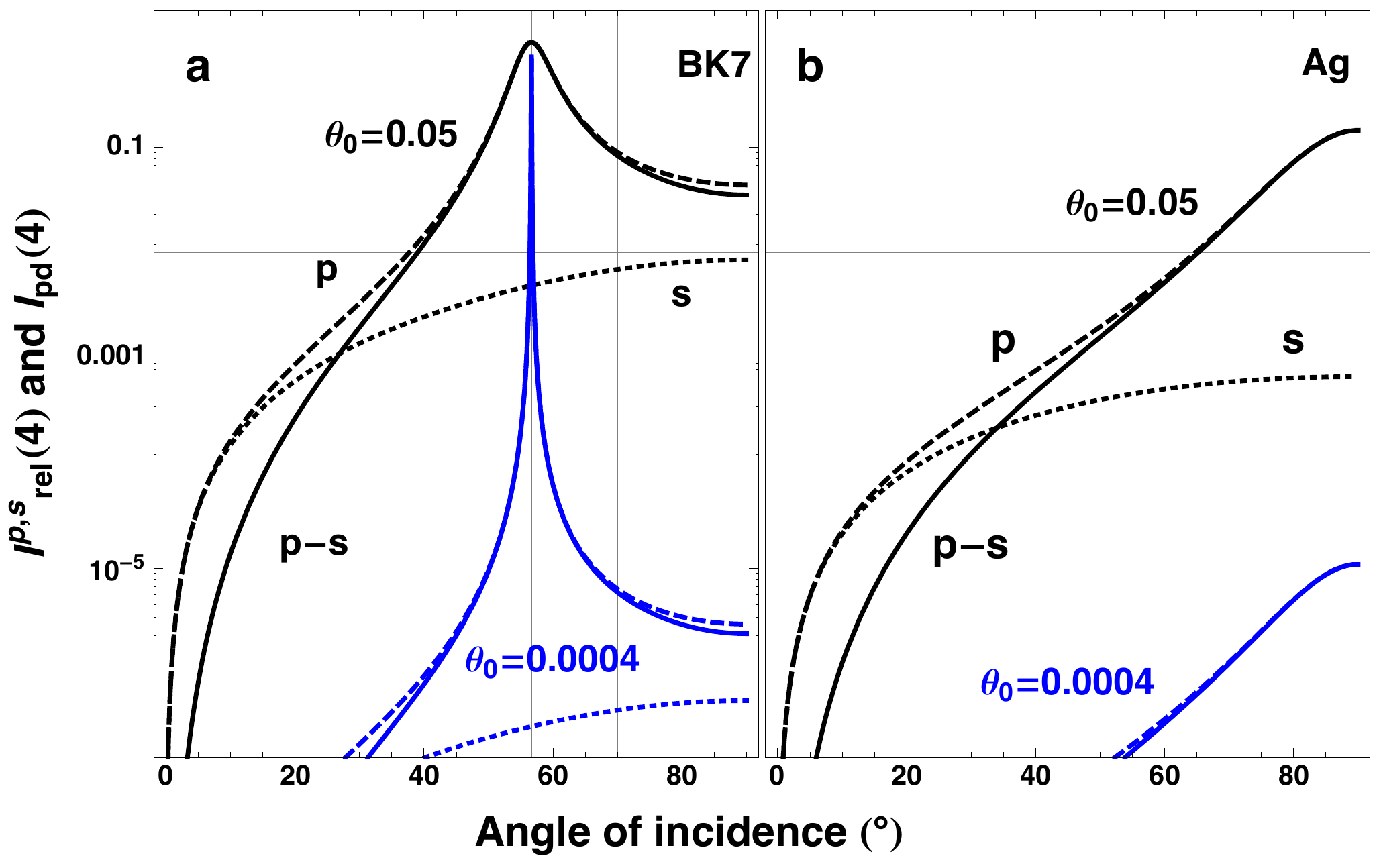}\caption{\label{fig:resth2}The relative $I_{rel}^{p,s}(\ell=4)$ and polarization-differential
$I_{pd}(\ell=4)$ intensity of the OAM sidebands (theory, logarithmic
plots) for external reflection. (a) shows the case for an air--BK7
interface ($n=1.52$), and (b) for a silver mirror; gray curves for
$\theta_{0}=0.0004$ (``collimated'') and black curves for $\theta_{0}=0.05$.
We see that in case of the glass interface, the experimentally accessible
polarization-differential sideband intensity (continuous curve) is
dominated by $p$ polarization (dashed curve), since $s$ polarization
(dotted line) does not experience the Brewster resonance. The vertical
lines indicate the Brewster angle and the angle for which we show
measurements in Fig.~\ref{fig:resexp}.}
\end{figure}

In Fig.~\ref{fig:resth2}a we present calculations of the SFC sidebands
for $\ell=4$, for the full range of angles of incidence, for $s$
and $p$ polarization separately, as well as for the polarization-differential
case. For $p$ polarization a Brewster resonance occurs due to the
vanishing of the reflection \cite{merano2009} making its contribution
to the sidebands in most cases much larger than that of $s$-polarization;
i.e. the polarization-differential $I_{pd}(\ell)$ is a sensible measure.
For $\theta_{0}=0.0004$ this admixture is (far) below 0.01, except
in a very narrow ($0.5^{\circ}$) angular window centered at the Brewster
angle. 

It is interesting to compare our results for an air-glass interface
with other cases, such as an air-metal interface. Our theory is fully
adequate for this since the material properties enter only via the
refractive index (or dielectric constant) which is complex-valued
for a metal. Fig.~\ref{fig:resth2}b gives the SFC sidebands $I_{pd}(\ell=4)$
for a silver interface ($n_{Ag}=0.14+4\, i$ at $\lambda=635$~nm).
In the collimated case, the sideband intensity is now much smaller
($<10^{-5}$ at all angles of incidence) than for an air-glass interface,
basically since a metal is a much better reflector. Also, the Brewster
resonance is absent in this case. We expect similar results for a
dielectric Bragg mirror. It will be interesting to check these predictions
experimentally. 

In conclusion, we have introduced the concept of spatial Fresnel coefficients
(SFC) to describe transverse-mode dependent reflection of a light
beam. In the OAM basis we find that an OAM mode $\{\ell\}$ acquires
sidebands. The sidebands are due to first-order diffractive corrections
to geometric optics (GH and IF effects); these lead to mode-dependent
displacement and thus to coupling to \textquotedblleft{}neighboring\textquotedblright{}
OAM modes. We find that these effects scale with the angular spread
of the beam. Only for well collimated beams they are small, which
bodes well for the use of folding mirrors in optical set-ups in the
laboratory and for mirror-assisted free-space OAM communication; contrarily,
the polarization degree of freedom can be severely complicated by
intervening mirrors. 

The OAM sidebands are enhanced for already moderately focussed beams.
We have used this case to validate our theory, by measuring the sidebands
occurring during external reflection by an air-glass interface, still
staying within the paraxial approximation. Practically, this case
could be encountered in the laboratory when using a beam splitter
to pick off a weak copy of a focused OAM light beam. The sidebands
disappear only in the limit of an ideal mirror with an infinite refractive
index step.

An intriguing question is, can we undo the SFC-induced mode coupling?
The reflective scattering process described by the matrix $c_{\ell,\ell'}^{\lambda}$
is reversible, so that in principle superposition input states can
be recovered with unity fidelity. But how can this be achieved experimentally?
An ordinary mirror cannot do this since it simply adds to the diffractive
beam shifts (GH and IF). We thus need an optical device that cancels
intrinsic beam shifts; possibly a negative-index metamaterial \cite{berman2002,lakhtakia2003,dolling2007}
or photonic crystal \cite{he2006} could achieve this task. 

\begin{acknowledgments}
We acknowledge fruitful discussions with M. J. A. de Dood, J. B. Götte,
and G. Nienhuis, and financial support by NWO and the EU STREP program
255914 (PHORBITECH).

\begin{center}
BIBLIOGRAPHY
\par\end{center}
\end{acknowledgments}
\bibliographystyle{apsrev4-1}
\bibliography{bibliography}

\begin{thebibliography}{31}%
\makeatletter
\providecommand \@ifxundefined [1]{%
 \@ifx{#1\undefined}
}%
\providecommand \@ifnum [1]{%
 \ifnum #1\expandafter \@firstoftwo
 \else \expandafter \@secondoftwo
 \fi
}%
\providecommand \@ifx [1]{%
 \ifx #1\expandafter \@firstoftwo
 \else \expandafter \@secondoftwo
 \fi
}%
\providecommand \natexlab [1]{#1}%
\providecommand \enquote  [1]{``#1''}%
\providecommand \bibnamefont  [1]{#1}%
\providecommand \bibfnamefont [1]{#1}%
\providecommand \citenamefont [1]{#1}%
\providecommand \href@noop [0]{\@secondoftwo}%
\providecommand \href [0]{\begingroup \@sanitize@url \@href}%
\providecommand \@href[1]{\@@startlink{#1}\@@href}%
\providecommand \@@href[1]{\endgroup#1\@@endlink}%
\providecommand \@sanitize@url [0]{\catcode `\\12\catcode `\$12\catcode
  `\&12\catcode `\#12\catcode `\^12\catcode `\_12\catcode `\%12\relax}%
\providecommand \@@startlink[1]{}%
\providecommand \@@endlink[0]{}%
\providecommand \url  [0]{\begingroup\@sanitize@url \@url }%
\providecommand \@url [1]{\endgroup\@href {#1}{\urlprefix }}%
\providecommand \urlprefix  [0]{URL }%
\providecommand \Eprint [0]{\href }%
\providecommand \doibase [0]{http://dx.doi.org/}%
\providecommand \selectlanguage [0]{\@gobble}%
\providecommand \bibinfo  [0]{\@secondoftwo}%
\providecommand \bibfield  [0]{\@secondoftwo}%
\providecommand \translation [1]{[#1]}%
\providecommand \BibitemOpen [0]{}%
\providecommand \bibitemStop [0]{}%
\providecommand \bibitemNoStop [0]{.\EOS\space}%
\providecommand \EOS [0]{\spacefactor3000\relax}%
\providecommand \BibitemShut  [1]{\csname bibitem#1\endcsname}%
\let\auto@bib@innerbib\@empty
\bibitem [{\citenamefont {Allen}\ \emph {et~al.}(1992)\citenamefont {Allen},
  \citenamefont {Beijersbergen}, \citenamefont {Spreeuw},\ and\ \citenamefont
  {Woerdman}}]{allen1992}%
  \BibitemOpen
  \bibfield  {author} {\bibinfo {author} {\bibfnamefont {L.}~\bibnamefont
  {Allen}}, \bibinfo {author} {\bibfnamefont {M.~W.}\ \bibnamefont
  {Beijersbergen}}, \bibinfo {author} {\bibfnamefont {R.~J.~C.}\ \bibnamefont
  {Spreeuw}}, \ and\ \bibinfo {author} {\bibfnamefont {J.~P.}\ \bibnamefont
  {Woerdman}},\ }\href {\doibase 10.1103/PhysRevA.45.8185} {\bibfield
  {journal} {\bibinfo  {journal} {Phys. Rev. A}\ }\textbf {\bibinfo {volume}
  {45}},\ \bibinfo {pages} {8185} (\bibinfo {year} {1992})}\BibitemShut
  {NoStop}%
\bibitem [{\citenamefont {Allen}\ \emph {et~al.}(1999)\citenamefont {Allen},
  \citenamefont {Padgett},\ and\ \citenamefont {Babiker}}]{allen1999}%
  \BibitemOpen
  \bibfield  {author} {\bibinfo {author} {\bibfnamefont {L.}~\bibnamefont
  {Allen}}, \bibinfo {author} {\bibfnamefont {M.}~\bibnamefont {Padgett}}, \
  and\ \bibinfo {author} {\bibfnamefont {M.}~\bibnamefont {Babiker}}\
  }(\bibinfo  {publisher} {Elsevier},\ \bibinfo {year} {1999})\ pp.\ \bibinfo
  {pages} {291 -- 372}\BibitemShut {NoStop}%
\bibitem [{\citenamefont {Mair}\ \emph {et~al.}(2001)\citenamefont {Mair},
  \citenamefont {Vaziri}, \citenamefont {Weihs},\ and\ \citenamefont
  {Zeilinger}}]{mair2001}%
  \BibitemOpen
  \bibfield  {author} {\bibinfo {author} {\bibfnamefont {A.}~\bibnamefont
  {Mair}}, \bibinfo {author} {\bibfnamefont {A.}~\bibnamefont {Vaziri}},
  \bibinfo {author} {\bibfnamefont {G.}~\bibnamefont {Weihs}}, \ and\ \bibinfo
  {author} {\bibfnamefont {A.}~\bibnamefont {Zeilinger}},\ }\href@noop {}
  {\bibfield  {journal} {\bibinfo  {journal} {Nature}\ }\textbf {\bibinfo
  {volume} {412}},\ \bibinfo {pages} {313} (\bibinfo {year}
  {2001})}\BibitemShut {NoStop}%
\bibitem [{\citenamefont {Molina-Terriza}\ \emph {et~al.}(2007)\citenamefont
  {Molina-Terriza}, \citenamefont {Torres},\ and\ \citenamefont
  {Torner}}]{molinaterriza2007}%
  \BibitemOpen
  \bibfield  {author} {\bibinfo {author} {\bibfnamefont {G.}~\bibnamefont
  {Molina-Terriza}}, \bibinfo {author} {\bibfnamefont {J.~P.}\ \bibnamefont
  {Torres}}, \ and\ \bibinfo {author} {\bibfnamefont {L.}~\bibnamefont
  {Torner}},\ }\href@noop {} {\bibfield  {journal} {\bibinfo  {journal} {Nat.
  Phys.}\ }\textbf {\bibinfo {volume} {3}},\ \bibinfo {pages} {305} (\bibinfo
  {year} {2007})}\BibitemShut {NoStop}%
\bibitem [{\citenamefont {{Dada}}\ \emph {et~al.}(2011)\citenamefont {{Dada}},
  \citenamefont {{Leach}}, \citenamefont {{Buller}}, \citenamefont
  {{Padgett}},\ and\ \citenamefont {{Andersson}}}]{dada2011}%
  \BibitemOpen
  \bibfield  {author} {\bibinfo {author} {\bibfnamefont {A.~C.}\ \bibnamefont
  {{Dada}}}, \bibinfo {author} {\bibfnamefont {J.}~\bibnamefont {{Leach}}},
  \bibinfo {author} {\bibfnamefont {G.~S.}\ \bibnamefont {{Buller}}}, \bibinfo
  {author} {\bibfnamefont {M.~J.}\ \bibnamefont {{Padgett}}}, \ and\ \bibinfo
  {author} {\bibfnamefont {E.}~\bibnamefont {{Andersson}}},\ }\href {\doibase
  10.1038/nphys1996} {\bibfield  {journal} {\bibinfo  {journal} {Nat. Phys.}\
  }\textbf {\bibinfo {volume} {7}},\ \bibinfo {pages} {677} (\bibinfo {year}
  {2011})}\BibitemShut {NoStop}%
\bibitem [{\citenamefont {Pors}\ \emph {et~al.}(2011)\citenamefont {Pors},
  \citenamefont {Miatto}, \citenamefont {'t~Hooft}, \citenamefont {Eliel},\
  and\ \citenamefont {Woerdman}}]{pors2011}%
  \BibitemOpen
  \bibfield  {author} {\bibinfo {author} {\bibfnamefont {B.-J.}\ \bibnamefont
  {Pors}}, \bibinfo {author} {\bibfnamefont {F.}~\bibnamefont {Miatto}},
  \bibinfo {author} {\bibfnamefont {G.~W.}\ \bibnamefont {'t~Hooft}}, \bibinfo
  {author} {\bibfnamefont {E.~R.}\ \bibnamefont {Eliel}}, \ and\ \bibinfo
  {author} {\bibfnamefont {J.~P.}\ \bibnamefont {Woerdman}},\ }\href@noop {}
  {\bibfield  {journal} {\bibinfo  {journal} {J. Opt.}\ }\textbf {\bibinfo
  {volume} {13}},\ \bibinfo {pages} {064008} (\bibinfo {year}
  {2011})}\BibitemShut {NoStop}%
\bibitem [{\citenamefont {Bechmann-Pasquinucci}\ and\ \citenamefont
  {Tittel}(2000)}]{bechmannpasquinucci2000}%
  \BibitemOpen
  \bibfield  {author} {\bibinfo {author} {\bibfnamefont {H.}~\bibnamefont
  {Bechmann-Pasquinucci}}\ and\ \bibinfo {author} {\bibfnamefont
  {W.}~\bibnamefont {Tittel}},\ }\href {\doibase 10.1103/PhysRevA.61.062308}
  {\bibfield  {journal} {\bibinfo  {journal} {Phys. Rev. A}\ }\textbf {\bibinfo
  {volume} {61}},\ \bibinfo {pages} {062308} (\bibinfo {year}
  {2000})}\BibitemShut {NoStop}%
\bibitem [{\citenamefont {Gibson}\ \emph {et~al.}(2004)\citenamefont {Gibson},
  \citenamefont {Courtial}, \citenamefont {Padgett}, \citenamefont {Vasnetsov},
  \citenamefont {Pas'ko}, \citenamefont {Barnett},\ and\ \citenamefont
  {Franke-Arnold}}]{gibson2004}%
  \BibitemOpen
  \bibfield  {author} {\bibinfo {author} {\bibfnamefont {G.}~\bibnamefont
  {Gibson}}, \bibinfo {author} {\bibfnamefont {J.}~\bibnamefont {Courtial}},
  \bibinfo {author} {\bibfnamefont {M.}~\bibnamefont {Padgett}}, \bibinfo
  {author} {\bibfnamefont {M.}~\bibnamefont {Vasnetsov}}, \bibinfo {author}
  {\bibfnamefont {V.}~\bibnamefont {Pas'ko}}, \bibinfo {author} {\bibfnamefont
  {S.}~\bibnamefont {Barnett}}, \ and\ \bibinfo {author} {\bibfnamefont
  {S.}~\bibnamefont {Franke-Arnold}},\ }\href@noop {} {\bibfield  {journal}
  {\bibinfo  {journal} {Opt. Express}\ }\textbf {\bibinfo {volume} {12}},\
  \bibinfo {pages} {5448} (\bibinfo {year} {2004})}\BibitemShut {NoStop}%
\bibitem [{\citenamefont {{Goos}}\ and\ \citenamefont
  {{H{\"a}nchen}}(1947)}]{goos1947}%
  \BibitemOpen
  \bibfield  {author} {\bibinfo {author} {\bibfnamefont {F.}~\bibnamefont
  {{Goos}}}\ and\ \bibinfo {author} {\bibfnamefont {H.}~\bibnamefont
  {{H{\"a}nchen}}},\ }\href {\doibase 10.1002/andp.19474360704} {\bibfield
  {journal} {\bibinfo  {journal} {Ann. Phys.}\ }\textbf {\bibinfo {volume}
  {436}},\ \bibinfo {pages} {333} (\bibinfo {year} {1947})}\BibitemShut
  {NoStop}%
\bibitem [{\citenamefont {Fedorov}(1955)}]{fedorov1955}%
  \BibitemOpen
  \bibfield  {author} {\bibinfo {author} {\bibfnamefont {F.~I.}\ \bibnamefont
  {Fedorov}},\ }\href@noop {} {\bibfield  {journal} {\bibinfo  {journal} {Dokl.
  Akad. Nauk SSSR}\ }\textbf {\bibinfo {volume} {105}},\ \bibinfo {pages} {465}
  (\bibinfo {year} {1955})}\BibitemShut {NoStop}%
\bibitem [{\citenamefont {Imbert}(1972)}]{imbert1972}%
  \BibitemOpen
  \bibfield  {author} {\bibinfo {author} {\bibfnamefont {C.}~\bibnamefont
  {Imbert}},\ }\href {\doibase 10.1103/PhysRevD.5.787} {\bibfield  {journal}
  {\bibinfo  {journal} {Phys. Rev. D}\ }\textbf {\bibinfo {volume} {5}},\
  \bibinfo {pages} {787} (\bibinfo {year} {1972})}\BibitemShut {NoStop}%
\bibitem [{\citenamefont {Bliokh}\ and\ \citenamefont
  {Bliokh}(2006)}]{bliokhprl2006}%
  \BibitemOpen
  \bibfield  {author} {\bibinfo {author} {\bibfnamefont {K.~Y.}\ \bibnamefont
  {Bliokh}}\ and\ \bibinfo {author} {\bibfnamefont {Y.~P.}\ \bibnamefont
  {Bliokh}},\ }\href {\doibase 10.1103/PhysRevLett.96.073903} {\bibfield
  {journal} {\bibinfo  {journal} {Phys. Rev. Lett.}\ }\textbf {\bibinfo
  {volume} {96}},\ \bibinfo {eid} {073903} (\bibinfo {year}
  {2006})}\BibitemShut {NoStop}%
\bibitem [{\citenamefont {Hosten}\ and\ \citenamefont
  {Kwiat}(2008)}]{hosten2008}%
  \BibitemOpen
  \bibfield  {author} {\bibinfo {author} {\bibfnamefont {O.}~\bibnamefont
  {Hosten}}\ and\ \bibinfo {author} {\bibfnamefont {P.}~\bibnamefont {Kwiat}},\
  }\href {\doibase 10.1126/science.1152697} {\bibfield  {journal} {\bibinfo
  {journal} {Science}\ }\textbf {\bibinfo {volume} {319}},\ \bibinfo {pages}
  {787} (\bibinfo {year} {2008})}\BibitemShut {NoStop}%
\bibitem [{\citenamefont {Aiello}\ and\ \citenamefont
  {Woerdman}(2008)}]{aiellobp2008}%
  \BibitemOpen
  \bibfield  {author} {\bibinfo {author} {\bibfnamefont {A.}~\bibnamefont
  {Aiello}}\ and\ \bibinfo {author} {\bibfnamefont {J.~P.}\ \bibnamefont
  {Woerdman}},\ }\href@noop {} {\bibfield  {journal} {\bibinfo  {journal} {Opt.
  Lett.}\ }\textbf {\bibinfo {volume} {33}},\ \bibinfo {pages} {1437} (\bibinfo
  {year} {2008})}\BibitemShut {NoStop}%
\bibitem [{\citenamefont {Schwefel}\ \emph {et~al.}(2008)\citenamefont
  {Schwefel}, \citenamefont {K\"{o}hler}, \citenamefont {Lu}, \citenamefont
  {Fan},\ and\ \citenamefont {Wang}}]{schwefel2008}%
  \BibitemOpen
  \bibfield  {author} {\bibinfo {author} {\bibfnamefont {H.~G.~L.}\
  \bibnamefont {Schwefel}}, \bibinfo {author} {\bibfnamefont {W.}~\bibnamefont
  {K\"{o}hler}}, \bibinfo {author} {\bibfnamefont {Z.~H.}\ \bibnamefont {Lu}},
  \bibinfo {author} {\bibfnamefont {J.}~\bibnamefont {Fan}}, \ and\ \bibinfo
  {author} {\bibfnamefont {L.~J.}\ \bibnamefont {Wang}},\ }\href@noop {}
  {\bibfield  {journal} {\bibinfo  {journal} {Opt. Lett.}\ }\textbf {\bibinfo
  {volume} {33}},\ \bibinfo {pages} {794} (\bibinfo {year} {2008})}\BibitemShut
  {NoStop}%
\bibitem [{\citenamefont {Gilles}\ \emph {et~al.}(2002)\citenamefont {Gilles},
  \citenamefont {Girard},\ and\ \citenamefont {Hamel}}]{gilles2002}%
  \BibitemOpen
  \bibfield  {author} {\bibinfo {author} {\bibfnamefont {H.}~\bibnamefont
  {Gilles}}, \bibinfo {author} {\bibfnamefont {S.}~\bibnamefont {Girard}}, \
  and\ \bibinfo {author} {\bibfnamefont {J.}~\bibnamefont {Hamel}},\
  }\href@noop {} {\bibfield  {journal} {\bibinfo  {journal} {Opt. Lett.}\
  }\textbf {\bibinfo {volume} {27}},\ \bibinfo {pages} {1421} (\bibinfo {year}
  {2002})}\BibitemShut {NoStop}%
\bibitem [{\citenamefont {Bliokh}\ and\ \citenamefont
  {Bliokh}(2007)}]{bliokhpre2007}%
  \BibitemOpen
  \bibfield  {author} {\bibinfo {author} {\bibfnamefont {K.~Y.}\ \bibnamefont
  {Bliokh}}\ and\ \bibinfo {author} {\bibfnamefont {Y.~P.}\ \bibnamefont
  {Bliokh}},\ }\href {\doibase 10.1103/PhysRevE.75.066609} {\bibfield
  {journal} {\bibinfo  {journal} {Phys. Rev. E}\ }\textbf {\bibinfo {volume}
  {75}},\ \bibinfo {eid} {066609} (\bibinfo {year} {2007})}\BibitemShut
  {NoStop}%
\bibitem [{\citenamefont {Vasnetsov}\ \emph {et~al.}(2005)\citenamefont
  {Vasnetsov}, \citenamefont {Pas'ko},\ and\ \citenamefont
  {Soskin}}]{vasnetsov2005}%
  \BibitemOpen
  \bibfield  {author} {\bibinfo {author} {\bibfnamefont {M.~V.}\ \bibnamefont
  {Vasnetsov}}, \bibinfo {author} {\bibfnamefont {V.~A.}\ \bibnamefont
  {Pas'ko}}, \ and\ \bibinfo {author} {\bibfnamefont {M.~S.}\ \bibnamefont
  {Soskin}},\ }\href@noop {} {\bibfield  {journal} {\bibinfo  {journal} {New J.
  Phys.}\ }\textbf {\bibinfo {volume} {7}},\ \bibinfo {pages} {46} (\bibinfo
  {year} {2005})}\BibitemShut {NoStop}%
\bibitem [{\citenamefont {Zambrini}\ and\ \citenamefont
  {Barnett}(2006)}]{zambrini2006}%
  \BibitemOpen
  \bibfield  {author} {\bibinfo {author} {\bibfnamefont {R.}~\bibnamefont
  {Zambrini}}\ and\ \bibinfo {author} {\bibfnamefont {S.~M.}\ \bibnamefont
  {Barnett}},\ }\href {\doibase 10.1103/PhysRevLett.96.113901} {\bibfield
  {journal} {\bibinfo  {journal} {Phys. Rev. Lett.}\ }\textbf {\bibinfo
  {volume} {96}},\ \bibinfo {pages} {113901} (\bibinfo {year}
  {2006})}\BibitemShut {NoStop}%
\bibitem [{\citenamefont {Okuda}\ and\ \citenamefont
  {Sasada}(2006)}]{okuda2006}%
  \BibitemOpen
  \bibfield  {author} {\bibinfo {author} {\bibfnamefont {H.}~\bibnamefont
  {Okuda}}\ and\ \bibinfo {author} {\bibfnamefont {H.}~\bibnamefont {Sasada}},\
  }\href {\doibase 10.1364/OE.14.008393} {\bibfield  {journal} {\bibinfo
  {journal} {Opt. Express}\ }\textbf {\bibinfo {volume} {14}},\ \bibinfo
  {pages} {8393} (\bibinfo {year} {2006})}\BibitemShut {NoStop}%
\bibitem [{\citenamefont {Okuda}\ and\ \citenamefont
  {Sasada}(2008)}]{okuda2008}%
  \BibitemOpen
  \bibfield  {author} {\bibinfo {author} {\bibfnamefont {H.}~\bibnamefont
  {Okuda}}\ and\ \bibinfo {author} {\bibfnamefont {H.}~\bibnamefont {Sasada}},\
  }\href {\doibase 10.1364/JOSAA.25.000881} {\bibfield  {journal} {\bibinfo
  {journal} {J. Opt. Soc. Am. A}\ }\textbf {\bibinfo {volume} {25}},\ \bibinfo
  {pages} {881} (\bibinfo {year} {2008})}\BibitemShut {NoStop}%
\bibitem [{\citenamefont {Aiello}\ and\ \citenamefont
  {Woerdman}(2004)}]{aiello2004b}%
  \BibitemOpen
  \bibfield  {author} {\bibinfo {author} {\bibfnamefont {A.}~\bibnamefont
  {Aiello}}\ and\ \bibinfo {author} {\bibfnamefont {J.~P.}\ \bibnamefont
  {Woerdman}},\ }\href {\doibase 10.1103/PhysRevA.70.023808} {\bibfield
  {journal} {\bibinfo  {journal} {Phys. Rev. A}\ }\textbf {\bibinfo {volume}
  {70}},\ \bibinfo {pages} {023808} (\bibinfo {year} {2004})}\BibitemShut
  {NoStop}%
\bibitem [{\citenamefont {Dennis}\ and\ \citenamefont
  {G{\"o}tte}(2012)}]{dennis2012}%
  \BibitemOpen
  \bibfield  {author} {\bibinfo {author} {\bibfnamefont {M.~R.}\ \bibnamefont
  {Dennis}}\ and\ \bibinfo {author} {\bibfnamefont {J.~B.}\ \bibnamefont
  {G{\"o}tte}},\ }\href@noop {} {\bibfield  {journal} {\bibinfo  {journal}
  {pre-print}\ } (\bibinfo {year} {2012})},\ \Eprint
  {http://arxiv.org/abs/1204.0327} {arXiv:1204.0327} \BibitemShut {NoStop}%
\bibitem [{\citenamefont {Bliokh}\ \emph {et~al.}(2009)\citenamefont {Bliokh},
  \citenamefont {Shadrivov},\ and\ \citenamefont {Kivshar}}]{bliokhvortex2009}%
  \BibitemOpen
  \bibfield  {author} {\bibinfo {author} {\bibfnamefont {K.~Y.}\ \bibnamefont
  {Bliokh}}, \bibinfo {author} {\bibfnamefont {I.~V.}\ \bibnamefont
  {Shadrivov}}, \ and\ \bibinfo {author} {\bibfnamefont {Y.~S.}\ \bibnamefont
  {Kivshar}},\ }\href@noop {} {\bibfield  {journal} {\bibinfo  {journal} {Opt.
  Lett.}\ }\textbf {\bibinfo {volume} {34}},\ \bibinfo {pages} {389} (\bibinfo
  {year} {2009})}\BibitemShut {NoStop}%
\bibitem [{\citenamefont {Merano}\ \emph {et~al.}(2010)\citenamefont {Merano},
  \citenamefont {Hermosa}, \citenamefont {Woerdman},\ and\ \citenamefont
  {Aiello}}]{merano2010}%
  \BibitemOpen
  \bibfield  {author} {\bibinfo {author} {\bibfnamefont {M.}~\bibnamefont
  {Merano}}, \bibinfo {author} {\bibfnamefont {N.}~\bibnamefont {Hermosa}},
  \bibinfo {author} {\bibfnamefont {J.~P.}\ \bibnamefont {Woerdman}}, \ and\
  \bibinfo {author} {\bibfnamefont {A.}~\bibnamefont {Aiello}},\ }\href
  {\doibase 10.1103/PhysRevA.82.023817} {\bibfield  {journal} {\bibinfo
  {journal} {Phys. Rev. A}\ }\textbf {\bibinfo {volume} {82}},\ \bibinfo
  {pages} {023817} (\bibinfo {year} {2010})}\BibitemShut {NoStop}%
\bibitem [{\citenamefont {Aiello}(2012)}]{aiello2012}%
  \BibitemOpen
  \bibfield  {author} {\bibinfo {author} {\bibfnamefont {A.}~\bibnamefont
  {Aiello}},\ }\href@noop {} {\bibfield  {journal} {\bibinfo  {journal} {New J.
  Phys.}\ }\textbf {\bibinfo {volume} {14}},\ \bibinfo {pages} {013058}
  (\bibinfo {year} {2012})}\BibitemShut {NoStop}%
\bibitem [{\citenamefont {Merano}\ \emph {et~al.}(2009)\citenamefont {Merano},
  \citenamefont {Aiello}, \citenamefont {van Exter},\ and\ \citenamefont
  {Woerdman}}]{merano2009}%
  \BibitemOpen
  \bibfield  {author} {\bibinfo {author} {\bibfnamefont {M.}~\bibnamefont
  {Merano}}, \bibinfo {author} {\bibfnamefont {A.}~\bibnamefont {Aiello}},
  \bibinfo {author} {\bibfnamefont {M.~P.}\ \bibnamefont {van Exter}}, \ and\
  \bibinfo {author} {\bibfnamefont {J.~P.}\ \bibnamefont {Woerdman}},\
  }\href@noop {} {\bibfield  {journal} {\bibinfo  {journal} {Nat. Photon.}\
  }\textbf {\bibinfo {volume} {3}},\ \bibinfo {pages} {337} (\bibinfo {year}
  {2009})}\BibitemShut {NoStop}%
\bibitem [{\citenamefont {Berman}(2002)}]{berman2002}%
  \BibitemOpen
  \bibfield  {author} {\bibinfo {author} {\bibfnamefont {P.~R.}\ \bibnamefont
  {Berman}},\ }\href {\doibase 10.1103/PhysRevE.66.067603} {\bibfield
  {journal} {\bibinfo  {journal} {Phys. Rev. E}\ }\textbf {\bibinfo {volume}
  {66}},\ \bibinfo {pages} {067603} (\bibinfo {year} {2002})}\BibitemShut
  {NoStop}%
\bibitem [{\citenamefont {Lakhtakia}(2003)}]{lakhtakia2003}%
  \BibitemOpen
  \bibfield  {author} {\bibinfo {author} {\bibfnamefont {A.}~\bibnamefont
  {Lakhtakia}},\ }\href@noop {} {\bibfield  {journal} {\bibinfo  {journal}
  {Electromagnetics}\ }\textbf {\bibinfo {volume} {23}},\ \bibinfo {pages} {71}
  (\bibinfo {year} {2003})}\BibitemShut {NoStop}%
\bibitem [{\citenamefont {Dolling}\ \emph {et~al.}(2007)\citenamefont
  {Dolling}, \citenamefont {Klein}, \citenamefont {Wegener}, \citenamefont
  {Sch{\"a}dle}, \citenamefont {Kettner}, \citenamefont {Burger},\ and\
  \citenamefont {Linden}}]{dolling2007}%
  \BibitemOpen
  \bibfield  {author} {\bibinfo {author} {\bibfnamefont {G.}~\bibnamefont
  {Dolling}}, \bibinfo {author} {\bibfnamefont {M.~W.}\ \bibnamefont {Klein}},
  \bibinfo {author} {\bibfnamefont {M.}~\bibnamefont {Wegener}}, \bibinfo
  {author} {\bibfnamefont {A.}~\bibnamefont {Sch{\"a}dle}}, \bibinfo {author}
  {\bibfnamefont {B.}~\bibnamefont {Kettner}}, \bibinfo {author} {\bibfnamefont
  {S.}~\bibnamefont {Burger}}, \ and\ \bibinfo {author} {\bibfnamefont
  {S.}~\bibnamefont {Linden}},\ }\href {\doibase 10.1364/OE.15.014219}
  {\bibfield  {journal} {\bibinfo  {journal} {Optics Express}\ }\textbf
  {\bibinfo {volume} {15}},\ \bibinfo {pages} {14219} (\bibinfo {year}
  {2007})}\BibitemShut {NoStop}%
\bibitem [{\citenamefont {He}\ \emph {et~al.}(2006)\citenamefont {He},
  \citenamefont {Yi},\ and\ \citenamefont {He}}]{he2006}%
  \BibitemOpen
  \bibfield  {author} {\bibinfo {author} {\bibfnamefont {J.}~\bibnamefont
  {He}}, \bibinfo {author} {\bibfnamefont {J.}~\bibnamefont {Yi}}, \ and\
  \bibinfo {author} {\bibfnamefont {S.}~\bibnamefont {He}},\ }\href@noop {}
  {\bibfield  {journal} {\bibinfo  {journal} {Opt. Express}\ }\textbf {\bibinfo
  {volume} {14}},\ \bibinfo {pages} {3024} (\bibinfo {year}
  {2006})}\BibitemShut {NoStop}%
\end{thebibliography}%

\end{document}